\documentclass{emulateapj}
\usepackage{epsfig}
\usepackage{graphicx}

\shorttitle{Evidence of Magnetic Interactions from 3D Doppler Tomography}
\shortauthors{Richards, Agafonov \& Sharova}

\begin{document}

\title{New Evidence of Magnetic Interactions between Stars from 3D Doppler Tomography of Algol Binaries: $\beta$ Per and RS Vul}

\author{Mercedes T. Richards}
\affil{Department of Astronomy \& Astrophysics, Pennsylvania State University, 525 Davey Laboratory, University Park, PA, 16802, USA}
\email{mrichards@astro.psu.edu} 

\author{Michail I. Agafonov} 
\affil{Radiophysical Research Institute (NIRFI), 25/12a, Bolshaya Pecherskaya St., Nizhny Novgorod, 603950, Russia}
\email{agfn@nirfi.sci-nnov.ru} 

\and
\author{Olga I. Sharova}
\affil{Radiophysical Research Institute (NIRFI), 25/12a, Bolshaya Pecherskaya St., Nizhny Novgorod, 603950, Russia}
\email{shol@nirfi.sci-nnov.ru}

\begin{abstract} 

Time-resolved H$\alpha$ spectra of magnetically-active interacting binaries have been used to create 3D Doppler tomograms by means of the Radioastronomical Approach.  This is the first 3D reconstruction of $\beta$ Per, with RS Vul for comparison.  These 3D tomograms have revealed evidence of the mass transfer process (gas stream, circumprimary emission, localized region, absorption zone), as well as loop prominences and coronal mass ejections (CMEs) in $\beta$ Per and RS Vul that could not be discovered from 2D tomograms alone.  The gas stream in both binaries may have been deflected beyond the central plane by the mass loser’s magnetic field.  The stream was more elongated along the predicted trajectory in RS Vul than in $\beta$ Per, but not as pronounced as in U CrB (stream-state). The loop prominence reached maximum $V_z$ velocities of $\pm$155 km~s$^{-1}$ in RS Vul compared to $\pm$120 km~s$^{-1}$ in $\beta$ Per, while the CME reached a maximum $V_z$ velocity of +150 km~s$^{-1}$ in RS Vul and +100 km~s$^{-1}$ in $\beta$ Per.  The 3D tomograms show that the gas flows are not symmetric relative to the central plane and are not confined to that plane; a result confirmed by recent 15GHz VLBI radio images of $\beta$ Per.  Both the 3D H$\alpha$ tomography and the VLBI radio images support an earlier prediction of the superhump phenomenon in $\beta$ Per: that the gas between the stars is threaded with a magnetic field even though the hot B8V mass gaining star is not known to have a magnetic field.

\end{abstract}


\section{Introduction}

The evolution of interacting binaries occurs as a consequence of mass transfer between the stars, and this process has been studied using spectroscopic and photometric analyses, theoretical simulations, and imaging techniques. Specifically, the image reconstruction technique of Doppler tomography has provided images showing how the gas moves (or flows) between and around the stars.  Moreover, tomography of time-resolved spectra has been used to determine observational measurements of the mass transfer rate in these binaries and to confirm the rates derived with the aid of the {\sc{shellspec}} spectrum synthesis code \citep{budaj+richards04,milleretal07}.  This derived rate is equal to the initial mass loss rate at the onset of Roche Lobe overflow since the donor star will transfer gas to its companion until the initial mass ratio has been reversed \citep{iben91}.  Hence, tomographic images of interacting binaries at various evolutionary stages yield important clues about the life cycles of interacting binaries. The early to late evolutionary stages of interacting binaries have been studied through spectroscopic analysis, calculation of synthetic spectra, and with both 2D and 3D tomography (e.g., see \citealt{richards12}).  

Two-dimensional tomography has provided images of: (1) accretion disks around the mass gaining star in the long-period Algols, CVs, nova-likes, X-ray binaries, and gamma-ray binaries; (2) gas streams and gas flowing along magnetic field lines in the magnetic CVs; (3) a combination of Keplerian disks, gas streams flowing along the predicted gravitational path, shock regions where the stream and disk interact, chromospheres, and other magnetic structures in Algol binaries; (4) regions where the gas slows down after circling the mass gainer in direct-impact systems (e.g., Algol, $\beta$ Per); and (5) asymmetric accretion disks that maintain their asymmetry on long timescales (e.g., TT Hya). For a summary of the 2D tomography, see \citet{richards04}.

Three-dimensional Doppler tomography was introduced using the Radioastronomical Approach (RA) developed by \citet{agafonov04a,agafonov04b} and \citet{agafonov+sharova05a,agafonov+sharova05b}. A summary of the important factors in the realization of 3D tomography is given in \citet{agafonov+sharova12}. To date, the 3D RA technique has been applied to only four systems: three Algol-type eclipsing binaries (U CrB, RS Vul, and now $\beta$ Per) and the X-ray binary Cyg X-1. 
The first 3D tomograms for the entire class of interacting binaries were created by \citet{agafonovetal06} using the U CrB alternating Algol system, which displays disklike emission in the tomogram at some epochs and streamlike emission at other epochs.  The 3D tomograms of the disk and stream states of U CrB were described by \citet{agafonovetal09} and compared to RS Vul by \citet{richardsetal10}.  A prominent gas stream was also discovered in the 2D and 3D images of Cyg X-1 \citep{sharovaetal12}.

The 3D images illustrate the distribution of gas flows in the orbital plane and beyond that plane, and confirm the range and complexity of the emission sources in these binaries: from accretion disks and gas streams, to shock regions and magnetic structures. Additional information provided by 3D tomography includes evidence that (6) prominent and extensive gas flows exist beyond the central plane of the binary in the $z$ direction (e.g., U CrB, RS Vul); (7) the accretion disk may precess or may be tilted relative to the central plane (e.g., U CrB); and (8) loop prominences and coronal mass ejections associated with the magnetic field of the donor star also contribute to the gas flows (e.g., RS Vul).  These results were summarized by \citet{richards12}.

In this paper, we continue our exploration of 3D tomography by comparing two magnetically-active systems: $\beta$ Per (Algol) and RS Vul.  Specifically, we have 1) identified the features in the 3D images and related them to those seen in the 2D tomograms; 2) compared the 3D tomogram of $\beta$ Per with those of RS Vul derived earlier; and 3) provided a physical interpretation of these features.   The system parameters are described in Section 2, the 2D tomograms are reviewed in Section 3, the 3D tomograms are described in Section 4, a model for $\beta$ Per and RS Vul based on the 3D images is given in Section 5, and the conclusions are stated in Section 6.

\section{System Properties for $\beta$ Per and RS Vul} 

$\beta$ Per (Algol) is the prototype of the Algol-type binaries.  It is the brightest and nearest eclipsing binary (d=29 pc), and one of the most extensively studied objects in the sky at all wavelengths.  It contains a partially eclipsing binary, with a tertiary in a 1.86 year orbit about the binary (see \citealt{richards+albright99}).  The system consists of a B8V primary, a K2IV secondary, and an F1V tertiary \citep{morgan35,richards93}. $\beta$ Per is a short-period Algol ($P_{orb}$ = 2.8673 days; \citealt{hilletal71}) with a mass ratio, $q$=0.22 $\pm 0.03$ \citep{richards93} and an orbital inclination of 81.4$ \pm 0.2^\circ$ \citep{richardsetal88}.  Other system properties include: $M_p$ = 3.7$\pm 0.3$ $M_\odot$, $M_s$ = 0.81 $\pm 0.05$ $M_\odot$, $R_p$ = 2.90 $\pm 0.0.04$ $R_\odot$, $R_s$ = 3.5 $\pm 0.1$ $R_\odot$ \citep{tomkin+lambert78,richards93}, systemic velocity $V_o$= 3.8 km s{$^{-1}$} \citep{hilletal71,eaton75}, and velocity semi-amplitude $K_p$ = 44.0 $\pm 0.4$ km s{$^{-1}$} \citep{hilletal71,tomkin+lambert78}.  \citet{baronetal12} used observations from the CHARA interferometer to determine new masses, radii and orbital properties of $\beta$ Per that are consistent with the parameters derived earlier.

The properties of $\beta$ Per are compared with those of RS Vul in Table 1; see \citet{richardsetal10} for the RS Vul references.  One notable difference is that RS Vul contains a G1 secondary compared to the K2 mass loser in $\beta$ Per.  Figure 1 also shows the similarities between the Roche geometries of the two binaries.  Both binaries have similar mass ratios and the $R_p$/$R_s$ ratio is $\sim 0.8$ for both systems even though the stars in RS Vul are larger than those in Algol, hence these binaries are expected to have similar accretion structures.

Magnetic activity associated with Algol secondaries is well known (e.g., \citealt{richards+albright93}), and both $\beta$ Per and RS Vul contain late-type magnetically-active secondaries.
$\beta$ Per is a strong radio and X-ray source \citep{wade+hjellming72,whiteetal86,
sternetal92,sternetal95}, and its cool secondary star is the dominant source of the emission at those wavelengths, characteristic of gyrosynchrotron radiation \citep{vandenoordetal89,favataetal00}.  Radio emission from RS Vul was detected with a flux density of 0.26 mJy at 5 GHz \citep{umanaetal98} compared to flux densities as high as 0.3 mJy at 2.3 GHz and 1.2 mJy at 8.3 GHz for $\beta$ Per \citep{richardsetal03}.  RS Vul also has an X-ray luminosity of $2.0 \times 10^{30}$ erg s$^{-1}$ \citep{white+marshall83}, that is about half of Algol's luminosity.  So, the magnetic field of Algol may be more the active than that of RS Vul. Hence a direct comparison between the 3D tomography and the new radio image should be interesting.

\begin{deluxetable}{lll} 
\tablecolumns{3} 
\tablewidth{0pc} 
\tabletypesize{\scriptsize}
\tablecaption{Properties of $\beta$ Per and RS Vul} 
\tablehead{ 
{Property} & {$\beta$ Per} & {RS Vul} 
}
\startdata
Spectral types	    &  B8V+K2IV+F1V 		&  B5V+G1III  \\
$P_{orb}$ (days)    & 2.867315  		&  4.4776635 \\
mass ratio, $q$     & 0.22 $\pm 0.03$   	& 0.27 \\
$i$ ($^{\circ}$)      & 81.4$ \pm 0.2^\circ$  	& 78.7$ \pm 0.2^\circ$ \\
$M_p$  ($M_\odot$)  & 3.7$\pm 0.3$ 		& 6.59$\pm 0.15$  \\
$M_s$  ($M_\odot$)  & 0.81 $\pm 0.05$ 		& 1.76 $\pm 0.05$  \\
$R_p$  ($R_\odot$)  &  2.90 $\pm 0.0.04$ 	& 4.71 $\pm 0.48$  \\
$R_s$  ($R_\odot$)  & 3.5 $\pm 0.1$ 		& 5.84 $\pm 0.23$  \\
$V_o$ (km s{$^{-1}$})  &  3.8  			& -20.1   \\
$K_p$ (km s{$^{-1}$})  & 44.0 $\pm 0.4$ 	& 54.0 $\pm 1.0$ 
\enddata
\end{deluxetable}

\begin{figure*}[t]
\figurenum{1}
\center
\psfig{file=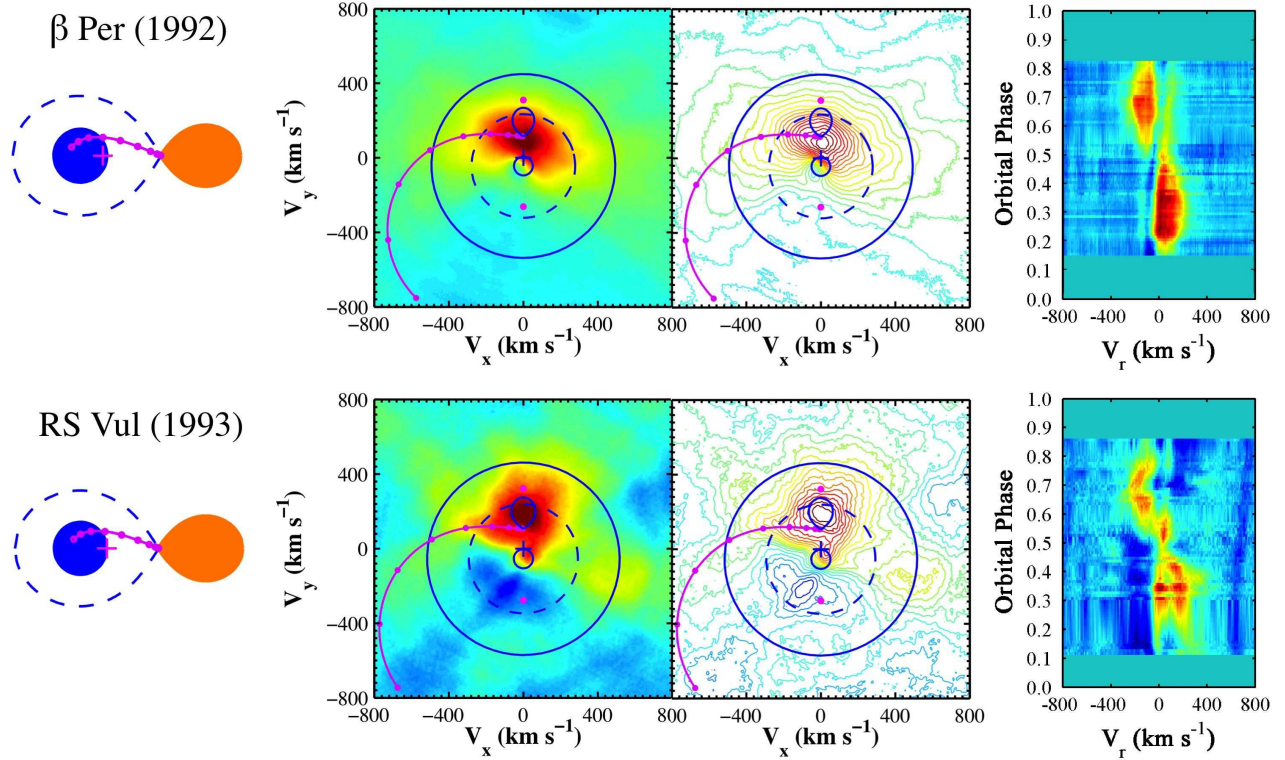,width=40pc}
\caption{Two-dimensional H$\alpha$ Doppler tomograms and contour maps of $\beta$ Per - 1992 (upper middle) and RS Vul - 1993 (lower middle), with Cartesian models of the two systems (left) and interpolated trailed spectrograms (right).  In the tomograms, the brightest sources are in deep red and the faintest in blue. Credit: \citet{richards12}. 
}
\label{f1}
\end{figure*}

\section{2D Doppler Images of $\beta$ Per and RS Vul} 

Two-dimensional Doppler tomograms of $\beta$ Per were created from one hundred and thirty-five time-resolved H$\alpha$ (6562.8 \AA) spectra of $\beta$ Per, with a dispersion of 0.093 \AA/pixel (4.3 km s{$^{-1}$}/pixel) or 6.9 \AA mm{$^{-1}$}, collected with the solar-stellar spectrograph on the 1.5m McMath-Pierce Telescope at the National Solar Observatory from 1992 October 6 - 21 (see \citealt{richards+albright99}). The spectra were obtained at closely spaced positions around the entire orbit of the binary, and all of the spectra used with the tomography procedure were obtained at phases outside of primary eclipse to exclude any contributions from the Rossiter-McLaughlin effect. Since the observed spectra are dominated by the spectrum of the B8 mass gaining star, difference spectra were calculated by subtracting theoretical photospheric spectra from the observed spectra (see \citealt{richards+albright99} for details). These difference spectra were used to calculate the tomograms of $\beta$ Per.

Figure 1 shows the 2D Doppler tomogram based on the 1992 spectra of $\beta$ Per along with the 2D tomogram of RS Vul (from \citealt{richards01}) based on H$\alpha$ spectra collected in two observing runs over 12 days from 1993 Apr 30 - May 31. These 2D images are later compared with the 3D Doppler tomograms based on the same data.  In the tomograms, the solid trajectory is the gravitational free-fall path of the gas stream; and the circles along this trajectory are marked at intervals of a tenth of the distance from the L$_1$ point to the distance of closest approach to the mass gainer.  The largest solid circle and the smaller dashed circle mark the inner and outer edge of a Keplerian disk, respectively; and the plus sign marks the center of mass of the binary, which is close to the stellar photosphere in the cases of $\beta$ Per and RS Vul.  The velocity images show the center of mass at $(V_x,V_y) = 0$, while the secondary star is placed at the velocity corresponding to its velocity semi-amplitude.  In the tomograms, the brightest sources are in deep red and the faintest in blue.

The first 2D tomogram of $\beta$ Per, based on 1976/77 spectra, was created by \citet{richardsetal96}.  That tomogram is similar to those derived from spectra collected in 1992 and again in 1994, more than 16 years later.  The 1976/77 trailed spectra of $\beta$ Per displayed both single- and double-peaked emission whose orbital variations followed S-wave patterns.  The sources of emission identified in the 2D tomogram were, from strongest to weakest: (1) a gas stream: the source of the single-peaked emission with velocities near the $L_1$ point; (2) a localized region: concentrated in the region directly between the stars and along the line of centers; (3) a sub-Keplerian crescent-shaped disk-like structure: the source of the double-peaked emission; and (4) an emission source on the velocity of the secondary star provided evidence of magnetic activity associated with that star. An independent S-wave analysis performed by \citet{richardsetal96} found that the source of the single-peaked emission was coincident with the part of the gas stream near the $L_1$ point, while  the source of the double-peaked emission was coincident with the concentrated Localized Region found in the 2D image. 

\section{The 3D Tomography Image Reconstruction}

The extension from 2D to 3D tomography has provided unprecedented views of the gas flows beyond the central plane in which the stars orbit. In this work, we used the Radioastronomical Approach of \citet{agafonov04a,agafonov04b} and \citet{agafonov+sharova05a,agafonov+sharova05b} that was developed further by application to four binaries (see \citealt{agafonovetal06}, \citealt{agafonovetal09}, and \citealt{richardsetal10}). A summary of the important factors in the realization of 3D tomography is given by \citet{agafonov+sharova12}. 

\begin{figure*}[t]
\figurenum{2}
\center
\epsscale{0.9}
\plotone{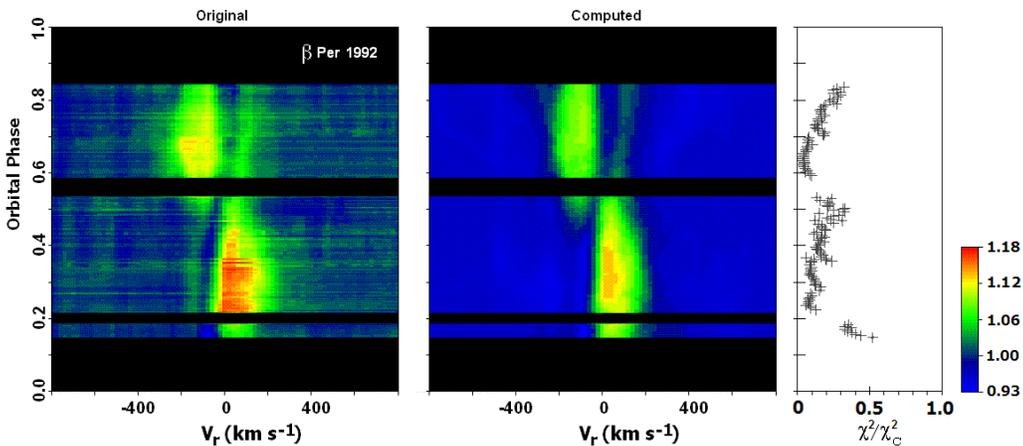}
\caption{Comparison between the original data (left frame) and the spectra computed from the reconstructed 3D Doppler map of $\beta$ Per (middle frame) in terms of the radial velocity, $V_r$ versus orbital phase.  The right frame displays the orbital phase variation of the relative chi-square statistic, $\chi^2/\chi_c^2$, where $\chi_c^2$ is the critical value corresponding to the 99\% confidence level.  The agreement between the observed and computed spectra is very good.
}
\label{f2}
\end{figure*} 

The basic assumptions and constraints of the 3D technique include \citep{richardsetal10}: (1) the spectra are assumed to be broadened primarily by Doppler motions; (2) spectra with high wavelength resolution and good resolution in orbital phase (or projections) are essential; (3) spectra dominated by emission lines are primarily used; (4) the assumption that the gas is optically thin is sufficient to create a good first-order view of the binary; (5) the orbital inclination influences the velocity resolution in the $V_z$ direction relative to those in the $V_x$ and $V_y$ directions, hence an adjustment is needed to compensate for the stretching effect of the Summarized Point Spread Function (SPSF) in the $V_z$ direction \citep{agafonovetal06}; (6) the transformation of the emission intensity from velocity space to coordinate Cartesian space is very complicated and it has not yet been solved even for 2D images. The RA technique creates 3D velocity maps of the gas intensity and it is an efficient method of processing the image in the case when only a limited number of spectra are available. 

\subsection{The 3D Tomogram of $\beta$ Per} 

The 3D tomogram of $\beta$ Per was constructed from the same H$\alpha$ spectra from 1992 October that were used to produce the 2D Doppler tomogram (see \citealt{richards01}). The 3D image was calculated with dimensions $(V_x,V_y,V_z)$ for values ranging from -800 to +800 km~s$^{-1}$, with intensities normalized to permit a comparison between slices. The velocity resolution of the reconstructed 3D image depends on the number of projections (i.e., number of spectra) and their distribution in orbital phase; while the orbital inclination influences the ratio of the resolutions in the $V_x$, $V_y$, and $V_z$ directions. The 4.3 km~s$^{-1}$ dispersion is high and certainly good enough to permit an optimal resolution of the 3D image since the dispersion plays a smaller role in setting the resolution of the image compared to the constraints set by the orbital inclination.  

Figure 2 shows the comparison between the observed spectra of $\beta$ Per and those computed from the reconstructed 3D tomograms.  This figure was used to examine the quality of the reconstructed 3D image produced by the RA method and it shows the trailed spectrograms in which the radial velocity, $V_r$ is plotted versus orbital phase.  The observed spectra are displayed in the left frames and the spectra computed from the 3D tomograms are displayed in the middle frames.  Two S-wave patterns are noticeable in the observed and computed spectra corresponding to a broad emission component and a narrow emission component.  The overall agreement between the observed and computed spectra is very good.   The quality of the fit is shown in the right frame of Figure 2, which displays the relative chi-square statistic, $\chi^2/\chi_c^2$, versus orbital phase.  Here, $\chi^2$ is normalized to the critical value of $\chi_c^2$, which corresponds to the largest acceptable value of $\chi^2$ at the 99\% confidence level and demonstrates the quality of the calculated values of $\chi^2$.  Figure 2 shows that $\chi^2$ was less than 30\% of the critical value at most orbital phases and it was only as high as 50\% of the critical value over a narrow phase range from 0.15 to 0.20, perhaps because the 3D code assumes that the gas is optically thin when it might be optically thick at these phases.  Since $\chi^2$ was much lower than the critical value the 3D reconstruction provides a good match to the data.

The 3D tomogram of $\beta$ Per is displayed in Figure 3 as fifteen slices in the ($V_x$,$V_y$) plane from $V_z$ = $-420$ km~s$^{-1}$ to $+420$ km~s$^{-1}$, at intervals of $60$ km~s$^{-1}$.  Two additional slices in the ($V_y$,$V_z$) direction are shown in the bottom frames.   Figure 3 shows that there are usually relatively small changes in the distribution of gas velocities between slices for a $V_z$ grid spacing of $60$ km~s$^{-1}$.  The most interesting representative slices in the 3D tomograms of $\beta$ Per and RS Vul are compared in Figure 4; for comparison, the full set of slices in the 3D tomogram of RS Vul can be found in \citet{richardsetal10}.  

\begin{figure*}
\figurenum{3}
\center
\epsscale{0.7}
\plotone{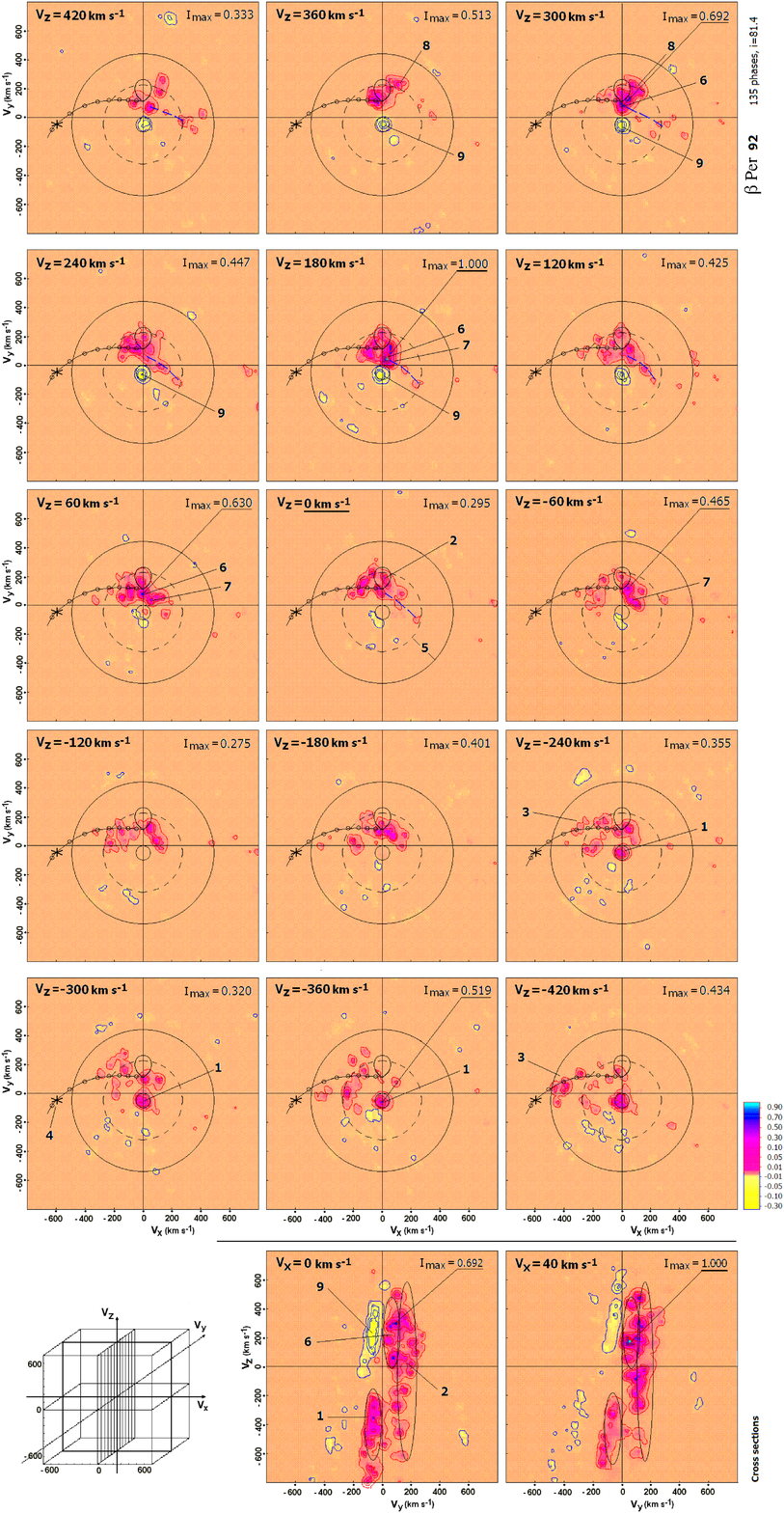}
\caption{Visualization of the ($V_x$,$V_y$) 2D slices in the 3D Doppler Tomogram of $\beta$ Per (1992) displayed symmetrically from $V_z$ = -420 km~s$^{-1}$  to $V_z$= +420 km~s$^{-1}$, in steps of 60 km~s$^{-1}$ in the $V_z$ direction.  Two additional images in the ($V_y$,$V_z$) direction are shown (bottom frames).  
}
\label{f3}
\end{figure*} 

Since the orbital inclination of $\beta$ Per is high ($i$ = 81.4$^\circ$), the velocity resolutions in the $V_x$ and $V_y$ directions will be better than the resolution in the $V_z$ direction. The resolution in the $V_z$ direction is the same as that in the other directions only for an inclination of 45$^\circ$, and degrades as the inclination increases toward 90$^\circ$. 
Consequently, the 3D tomogram of $\beta$ Per was restored with a resolution of 40 km~s$^{-1}$ in the $V_x$ and $V_y$ directions and 180 km~s$^{-1}$ in the $V_z$ direction, corresponding to a SPSF half power beam width (HPBW) of $40 \times 40 \times 180$ in units of km~s$^{-1}$.  In order to adjust for the artificial stretching effect of the SPSF, the $V_z$ velocities were scaled by a factor of 0.22 (= 40/180), corresponding to the ratio of the velocity resolutions in the $V_z$ direction compared to the other directions. This uniform adjustment is useful in illustrating the locations of the emission sources in the 3D image.  The adjusted $V_z$ values are given in the last column of Table 2.   

\begin{figure*}
\figurenum{4}
\center
\epsscale{1}
\plotone{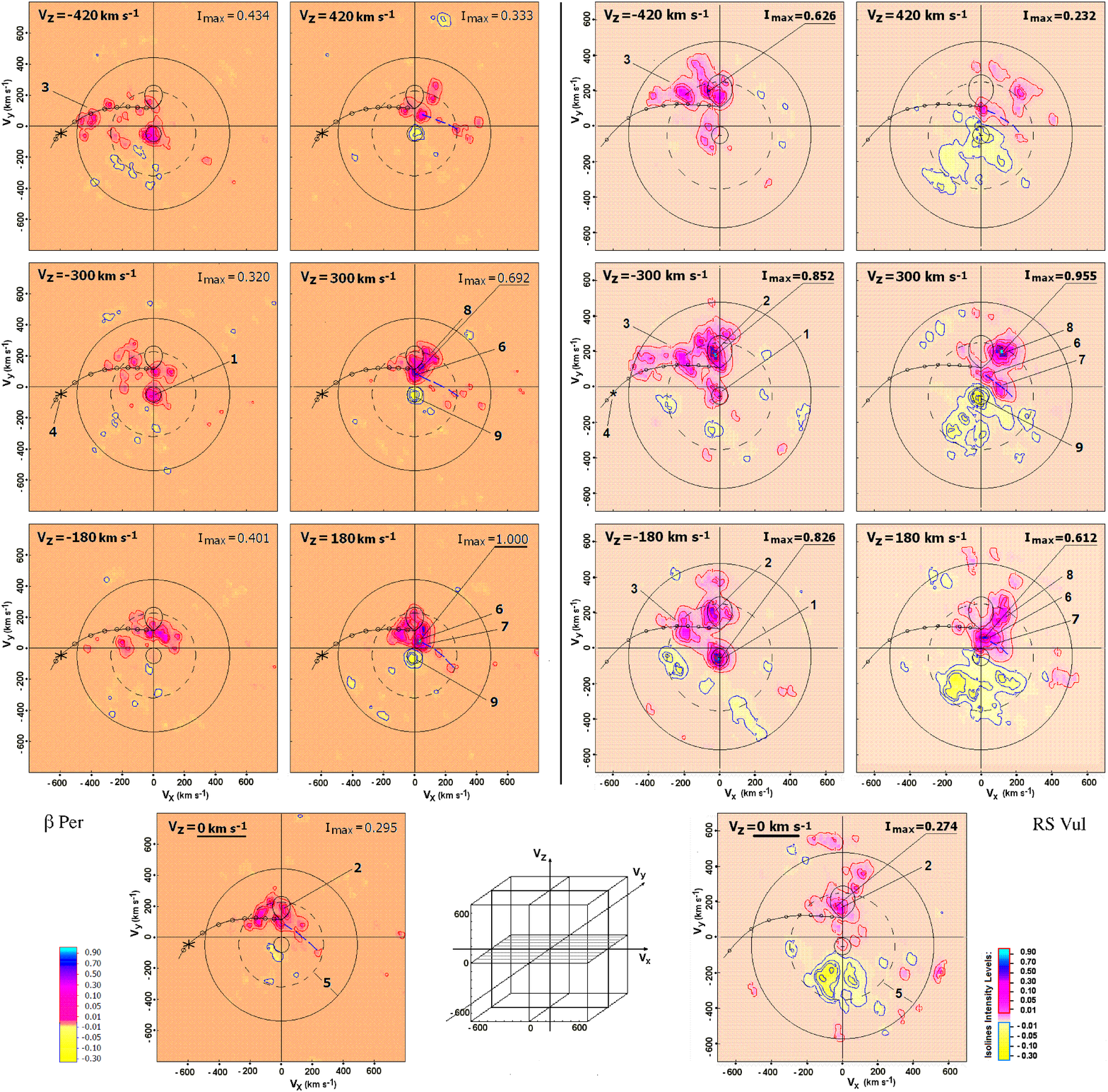}
\caption{A comparison of symmetric ($V_x$,$V_y$) slices along the $V_z$ direction in the 3D Doppler tomograms of $\beta$ Per and RS Vul. The blue dashed lines illustrate the trajectory of the gas that flowed around the mass gainer only to be slowed by impact with the incoming gas stream and the Localized Region (shown for both binaries).
}
\label{f4}
\end{figure*} 

\begin{deluxetable*}{clllll}
\tablecolumns{6} 
\tablewidth{0pc} 
\tabletypesize{\scriptsize}
\tablecaption{Characteristics and locations of prominent emission features in the 3D tomogram of $\beta$ Per} 
\tablehead{  
\colhead{} & \colhead{ } & \multicolumn{4}{c}{Location: central velocity or velocity range (km~s$^{-1}$)} 
} 
\startdata
{No.} & {Emission Feature} & {$V_x$} & {$V_y$}  & {$V_z$} & {$V_z(adjusted)$} \\
\hline 
1 & Circumprimary emission          & 0 (-60 to +60)                 & -60 (0 to -120)                & -400  (-200 to 600)   & -100 (-50 to -150) \\
2 & Emission on donor star (active regions)  & 0 (-70 to +70)                 & 180 (80 to 280)                &  0 (-500 to +500)     & 0 (-120 to +120)  \\
  & Emission near L1                & 0               		     & +110                           & -0 (-500 to +500)     & -0 (-120 to +120)  \\
3 & Gas stream flow                 & -520 to 0 {\tablenotemark{a}}  & 0 to 300 {\tablenotemark{a}} & -300 (-480 to +120) & -70 (-115 to +30) \\
4 & Star-stream impact region       & No source found                &                                &  		    & \\
5 & Locus of the accretion disk     & No source found                &                                &                     &  \\

6 & Localized Region (LR) -- Part 1 & 0 (-50 to +50)                 & +50 (0 to 100)                 & 200 (0 to 400)     & 50 (0 to 100)  \\
7 & Localized Region -- Part 2      & 70 (20 to 120)                 & +50 (-20 to 140)               & 60 (-180 to 240)    & 15 (-45 to 60) \\
8 & High velocity flow near donor  & 110 (50 to 170)                 & 220 (150 to 300)               & 340 (240 to 440)    & 80 (60 to 100) \\
9 & Absorption zone                 & 0 (-60 to +60)                 & -60 (0 to -120)                & 240 (80 to 460)       & 60 (20 to 115) \\
\enddata
\tablenotetext{a}{corresponds to the predicted ballistic trajectory of the gas stream.}
\end{deluxetable*}

Figure 5 illustrates how the SPSF affects the resolution of the image. Its pattern on the HPBW in the ($V_x$,$V_y$) plane (see left frame) leads to the equal resolution along the $V_x$ and $V_y$ axes. However, the image is also stretched approximately four times in the $V_z$ direction relative to the other directions (see the ($V_y$,$V_z$) plane in the right frame).  An example of the deconvolution to adjust for this artificial distortion is illustrated in Figure 6.  This figure shows cross sections of the 3D tomograms in the ($V_y$,$V_z$) plane for $V_x$=0 km~s$^{-1}$, before and after the adjustment  to the velocities in the $V_z$ direction.  A uniform scaling of the images by the factor of 0.22 provides a rough first approximation to the complex deconvolution that is needed to rectify the images.

\subsection{Description and Interpretation of Emission Features} 

The dynamical properties of the mass transfer process in $\beta$ Per were expected to be similar to those of RS Vul based on their Roche lobe geometries shown in Figure 1.  The 3D tomograms shown in Figures 3 and 4 confirm that the 3D reconstruction has reproduced the 2D tomograms for the case when $V_z$ = 0 km~s$^{-1}$.  Moreover Figure 4 shows some general similarities between the 3D images of these two binaries.  

\begin{figure}
\figurenum{5}
\center
\epsscale{1.1}
\plotone{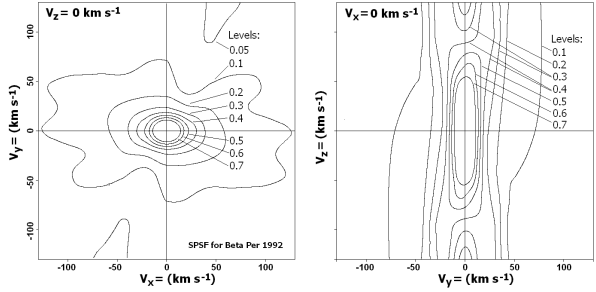}
\caption{Slices of the SPSF calculated for all the 3D viewing angles (inclination and orbital phases) of the $\beta$ Per binary.  The SPSF affects the resolution of the image in a symmetric way in the ($V_x$,$V_y$) plane (left frame) while artificially distorting the image along the $V_z$ direction in the ($V_y$,$V_z$) plane  (right frame).  This effect is strongly influenced by the inclination of the binary to the plane of the sky. 
}
\label{f5}
\end{figure} 

\begin{figure}
\figurenum{6}
\center
\epsscale{1.15}
\plotone{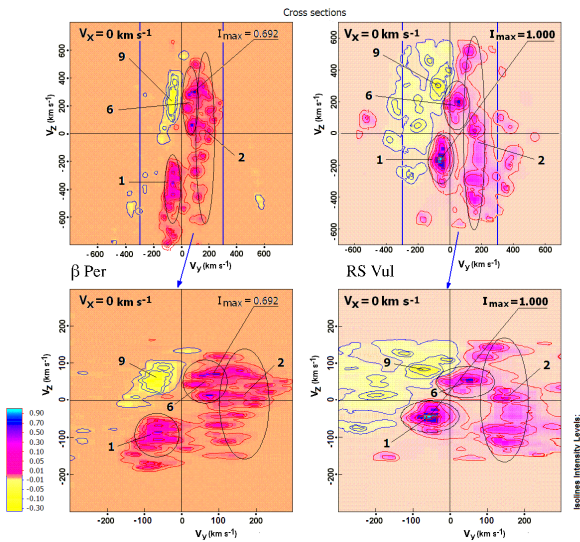}
\caption{Cross sections of the $\beta$ Per and RS Vul 3D Doppler tomograms in the ($V_y$,$V_z$) plane for $V_x$=0 km~s$^{-1}$. The top frames show the original images, and the bottom frames show the central parts of the same slices after an approximate deconvolution to adjust for the artificial stretching effect of the SPSF in the $V_z$ direction. The numbered features correspond to those in Figs. 3 and 4.
}
\label{f6}
\end{figure} 

Several main features and other weaker features were detected in the tomograms shown in Figures 3 and 4.  The numbering system for these features is given in Table 2 and is consistent with earlier papers in this series. The $V_z$ velocities, $V_z(adj)$, quoted below are the values after the adjustment for the stretching effect in this direction. 

The strongest sources in $\beta$ Per are listed below. \hfill\break
\noindent
{\bf Feature 1}:  An emission source associated with the primary star, the circumprimary emission, was detected with high negative $V_z(adj)$ velocities from -50 to -150 km~s$^{-1}$ in $\beta$ Per compared to 0 to +90 km~s$^{-1}$ in RS Vul.   This source provides evidence that the mass gainer was spun-up to super-synchronous values by the impact of the gas stream onto the stellar surface.  Moreover, the direction of the flow in the $V_z$ direction in $\beta$ Per was opposite to that in RS Vul.  In both binaries, the projected rotational velocity, $v\sin i$ is comparable to the synchronous rotational velocity of the mass gainer (50 - 60 km~s$^{-1}$), and both are lower than the maximum $V_z(adj)$ velocities associated with this feature (e.g., $v\sin i = 53 \pm 3$ km~s$^{-1}$ for $\beta$ Per; \citealt{rucinski79, tomkin+tan85}).   Since the $V_z(adj)$  velocities associated with Feature 1 are higher than the star's rotational velocity, it is reasonable to assume that the star may have been spun up, presumably by the impact of the gas stream onto the stellar photosphere.    Similar super-synchronous rotation velocities which have been measured directly from line profile analysis of direct-impact systems including $\beta$ Per (e.g., \citealt{mallama78a,mallama78b}).  Velocities of $\sim$500 km~s$^{-1}$ at the impact site were predicted by \citet{lubow+shu75} using semi-analytical ballistic calculations of direct-impact systems; such high velocities could produce supersonic turbulence and increase the angular momentum of the material around the mass gainer \citep{richards93}.

\vskip2pt
\noindent
{\bf Feature 2}: An emission source associated with active magnetic regions on the donor-star was detected with a wide range of $V_z(adj)$ velocities from -120 to +120 km~s$^{-1}$, equally distributed about the central plane.  In the ($V_x$,$V_y$) slices for both binaries, this feature has nearly the same velocity as the donor star.  However, in RS Vul, this emission displayed a wider $V_z$ range from -150 to +155 km~s$^{-1}$, suggesting higher magnetic activity than found for $\beta$ Per.  This behavior over a 305 km~s$^{-1}$ range in $V_z(adj)$ led to the discovery of a loop prominence in RS Vul along which gas is rising and falling since the footpoint in the ($V_x$,$V_y$) plane stayed the same while the $V_z$ velocity changed \citep{richardsetal10}.  The $V_z(adj)$ range of 240 km~s$^{-1}$ in $\beta$ Per suggests that yet another prominence has been found in Algol-type binaries. 

\vskip2pt
\noindent
{\bf Feature 3}:  An emission source associated with the gas stream along the predicted gravitational trajectory in the ($V_x$,$V_y$) plane was detected at $V_z(adj)$ velocities from -115 to +30 km~s$^{-1}$. This source reached the same high velocities found in RS Vul, but it was not as prominent (or bright) in $\beta$ Per as in RS Vul.  In both binaries, this source reached peak extension at high negative $V_z$ velocities, and not in the central plane ($V_z$ = 0 km~s$^{-1}$), suggesting that the gas stream may have been deflected away from the central ($V_x$,$V_y$) plane by the magnetic effects of the mass losing stars in both binaries.  

\vskip2pt
\noindent
{\bf Feature 4}: The star-stream impact region was not detected in either binary even though the gas stream flow reached the velocity associated with the surface of the mass gaining star in both cases.

\vskip2pt
\noindent
{\bf Feature 5}: Very little emission was found within the locus of a Keplerian accretion disk in Algol, and only a small amount of gas was detected in the case of RS Vul.

\vskip2pt
\noindent
{\bf Features 6, 7}: The Localized Region (LR) between the stars was detected in both binaries over a range of $V_z$ velocities.  This source represents the part of the gas flow that circles the star and interacts with the incoming gas stream.  The blue dashed lines in Figure 4 illustrate the trajectory of the gas that flowed around the mass gainer only to be slowed by impact with the incoming gas stream and the Localized Region.

\vskip2pt
\noindent
{\bf Feature 8}: The high velocity flow associated with the donor star reached $V_z(adj)$ velocities of +100 km~s$^{-1}$ in $\beta$ Per, compared to +150 km~s$^{-1}$ in RS Vul.  Relative to the general accretion flow, these large positive velocities suggest that gas was ejected from the donor star in a manner similar to a coronal mass ejection (CME), which can achieve velocities from hundreds of km~s$^{-1}$ to over 1000 km~s$^{-1}$ in extreme cases on the Sun.  This result suggests once again that the cool donor star was more active in RS Vul than in $\beta$ Per.

\vskip2pt
\noindent
{\bf Feature 9}: An absorption source associated with the velocity of the mass gaining star in the ($V_x$,$V_y$) plane was detected at positive $V_z(adj)$ velocities, from +20 to +115 km~s$^{-1}$.   In RS Vul, this feature had higher velocities in the ($V_x$,$V_y$) plane, over a wider range of velocities in that plane, and for $V_z(adj)$ velocities from -35 to +155 km~s$^{-1}$.  This source provides additional evidence that the mass gainer was spun-up by the impact of the gas stream onto the stellar surface and heated to temperatures beyond the optical regime. A similar effect was found for U Sge when the H$\alpha$ tomogram was compared with the uv tomogram \citep{kempner+richards99}.  The uv tomogram showed that the hotter gas was roughly in the same part of the tomogram as the absorption region near the accreting star.

\begin{figure*}
\vspace{-1cm}
\figurenum{7}
\center
\psfig{file=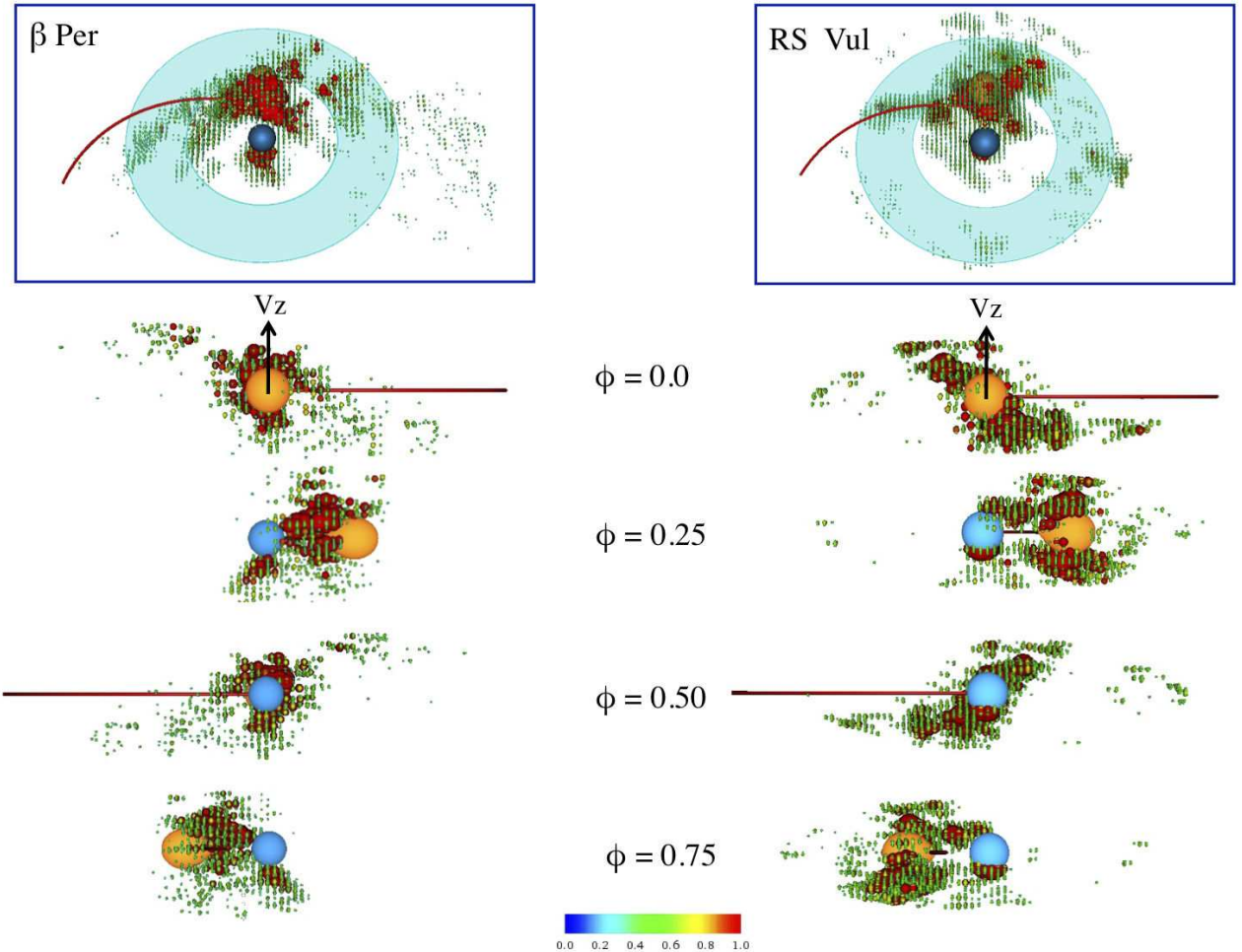,width=45pc}
\vspace{-1cm}
\caption{3D representations of the velocity distribution of emission sources relative to the $V_z$ axis for RS Vul and $\beta$ Per.  The top frames illustrate a tilted view of the velocity distribution of the gas beyond the central velocity plane of each binary.  The other 3D views are shown in order of orbital phase starting with $\phi$ = 0.0 (second row), $\phi$ = 0.25 (third row), $\phi$ = 0.5 (fourth row), and $\phi$ = 0.75 (last row).  The emission sources in the image have been scaled according to intensity, with the brightest sources shown as large red points and the fainter sources as small green points.
}
\label{f7}
\end{figure*}

\section{A Model for the Magnetically Active Binaries of $\beta$ Per and RS Vul}

In summary, the general features found in 2D slices of the 3D tomograms of the interacting binaries U CrB, RS Vul, $\beta$ Per, and  Cyg X-1 include: ({\bf 1}) circumprimary emission, ({\bf 2}) emission from active magnetic regions associated with the donor star, ({\bf 3}) flow along the predicted trajectory of the gas stream, ({\bf 4}) the star-stream impact site, ({\bf 5}) the predicted locus of a Keplerian accretion disk, ({\bf 6}) the localized region between stars, ({\bf 7}) a second localized region, ({\bf 8}) high velocity flow moving away from the donor star, and ({\bf 9}) the absorption zone  near the mass gaining star.  The characteristics of most of these sources found in $\beta$ Per were either identical or similar to those found for RS Vul (see \citealt{richardsetal10}). However,  the star-stream interaction ({\bf 4}) was absent in both binaries, and the Keplerian disk ({\bf 5}) was absent in $\beta$ Per and weak in RS Vul.  The absorption zone ({\bf 9}) in $\beta$ Per was found with low velocities near the mass gainer, but with a higher range of velocities in RS Vul.  Figure 7 shows the 3D representations of the velocity distribution of emission sources relative to the $V_z$ axis for both binaries. 

The emission sources associated with the cool magnetically active star (Features {\bf 2} and {\bf 8}) are of particular interest in this work.  Both $\beta$ Per and RS Vul displayed evidence of loop prominences ({\bf 2}) and coronal mass ejections ({\bf 8}).  Moreover, these flows are not symmetric relative to the central plane. Since the tomogram reveals the Doppler motion of the gas, then the detection of a smooth transition from positive $V_z$ to negative $V_z$ velocities relative to a fixed footpoint in the central ($V_x$,$V_y$) plane is evidence of a smooth motion toward and away from the observer in the $V_z$ direction beyond the central plane (see Table 2).  This behavior is characteristic of a loop prominence similar to those detected on the Sun.  Similarly, a smooth transition from low to high positive $V_z$ velocities relative to a fixed footpoint in the central ($V_x$,$V_y$) plane suggests motion away from the observer at high velocities.  This behavior is reminiscent of the mass loss associated with flares known as CMEs.  For both phemenona, RS Vul achieved higher $V_z$ velocities than $\beta$ Per.  Moreover, 2D Doppler tomograms based on spectra of $\beta$ Per collected from 1976 to 1994, showed that the image of $\beta$ Per remained almost the same over that 18-year period of time (e.g., \citealt{richardsetal96}).

Both of these phenomena have now been detected in $\beta$ Per and RS Vul with the help of 3D Doppler tomography.  This behavior was to be expected based on the radio and X-ray evidence that both systems should be magnetically active (See Section 2).  A 5.6-year continuous radio flare survey by \citet{richardsetal03} revealed that radio flares occur regularly every 48.9 $\pm$ 1.7 days on the cool mass-losing star in $\beta$ Per at both 2.3 GHz and 8.3 GHz. This cycle corresponds to a major flare every 17 orbital periods or roughly every 1.6 months.  Hence it should not be surprising that a CME was detected in a tomogram collected over a period of 16 days in October 1992.  RS Vul was too faint to be included in that survey, so its flaring period is still unknown.

The continuous radio survey of $\beta$ Per at 2.3 GHz and 8.3 GHz also yielded evidence of superhump-like behavior, suggesting that the gas stream from the cool star is threaded by that star's magnetic field \citep{retteretal05}.  This was the first detection of the superhump phenomenon in the radio regime and the first observation of superhumps in Algol systems.  Moreover, this result suggests that the gas in the vicinity of the mass gaining star (flowing along the path of a sub-Keplerian annulus) in $\beta$ Per is threaded with a magnetic field even though this B8V star is not known to have a magnetic field.   Instead, the magnetic field in the annulus originates from amplification of the seed field in the magnetized material transferred from the mass losing star. Since the ram pressure of the accretion flow is larger than the magnetic stress at the Lagrangian point, the ionized accreting material drags the field along \citep{retteretal05}. The field lines are wound around the mass gaining star through the gas stream; the field is  stretched and amplified, and at some stage reconnection can occur, causing flaring and acceleration of the particles.  The exception in the case of $\beta$ Per is that only one star in the system has a strong magnetic field.  In spite of the field reconnection in the gas around the mass gainer, the lack of radio emission near the B8V star could be explained by the Razin effect and/or plasma low-frequency cutoff when the electron density is high (e.g., \citealt{dulk85}).

A consequence of this superhump effect is that the circumprimary gas is expected to precess.   This is interesting since \citet{agafonovetal09} used 3D Doppler tomography to predict that the Keplerian accretion disk in another Algol-type binary, U CrB, does precess.  So, it is satisfying to discover that the superhump phenomenon is not restricted to the classical Keplerian accretion disks found in compact binaries.

The superhump result predicted by \citet{retteretal05} was confirmed five years later by \citet{petersonetal10} through a global very long baseline radio interferometer array (VLBI) study which produced a resolved 15 GHz radio image of $\beta$ Per based on observations collected in 2008.  This radio image displayed evidence of gyrosynchrotron filled-loops associated with the cool donor star, a magnetically-threaded gas stream between the stars, and asymmetric magnetic structures that were not confined to the orbital plane; however, no radio emission was directly associated with the mass gaining star. Figure 8 shows the comparison between the VLBI results and those derived from 3D Doppler tomography of $\beta$ Per.  The resemblance suggests that tomography is truly a powerful tool that provides information about the gas flows that is consistent with the radio image of the binary, at least in the case of $\beta$ Per.  Our 3D results confirm the \citet{petersonetal10} results, and also provide additional information about the interactions between the two stars beyond the orbital plane. 

\begin{figure*}
\figurenum{8}
\center
\psfig{file=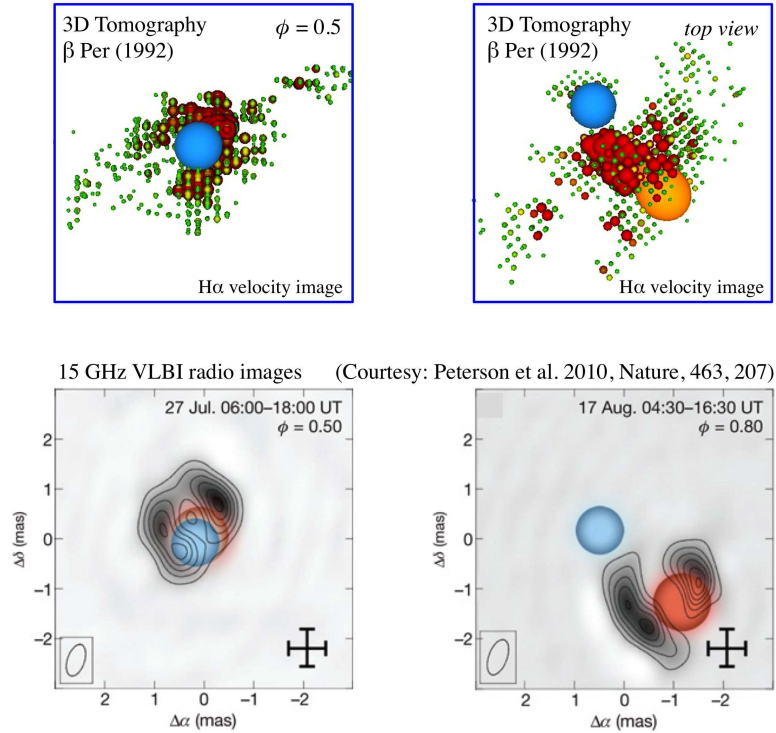,width=25pc}
\caption{Comparison between the 3D tomography H$\alpha$ images of $\beta$ Per (top frames) with the 15GHz VLBI radio images of the binary at two phases (lower frames) derived from \citet{petersonetal10}.  The top images are velocity images while the lower images are position images.  Magnetic activity can be detected both in the radio and H$\alpha$ spectra and these images display consistent information about the binary.
}
\label{f8}
\end{figure*} 

\section{Conclusions}

We have used time-resolved H$\alpha$ spectra of two magnetically-active interacting binaries to extract information about the three-dimensional velocity distribution of the gas flows in these systems.  This is the first 3D reconstruction of $\beta$ Per using the Radioastronomical Approach. Emission and absorption features detected in 2D tomograms have been confirmed in the 3D Doppler tomograms of $\beta$ Per and RS Vul.  The 3D images have revealed evidence of the mass transfer process, including the gas stream, circumprimary emission probably caused by the impact of the gas stream onto the surface of the hot mass gaining star, an absorption zone where gas has been heated to temperatures beyond the H$\alpha$ regime, and a localized region where the gas that circled the mass gainer has been slowed to sub-Keplerian velocities by impact with the incoming gas stream. The 3D images also display evidence of magnetic activity associated with the cool star, including evidence of loop prominences and coronal mass ejections that could not be discovered from the 2D tomograms alone.  

The extension from 2D to 3D images has provided a view of the gas flowing beyond the central plane defined by the orbital motions of the stars.  The 3D images of $\beta$ Per and RS Vul have once again revealed that the gas flowing beyond the central plane is substantial since the gas achieved $V_z(adj)$ velocities as high as $\pm$155 km~s$^{-1}$.  Specifically, the loop prominence reached maximum $V_z(adj)$ velocities of -150 to +155 km~s$^{-1}$ in RS Vul compared to -120 to +120 km~s$^{-1}$ in $\beta$ Per, and the CME reached a maximum $V_z(adj)$ velocity of +150 km~s$^{-1}$ in RS Vul and +100 km~s$^{-1}$ in $\beta$ Per.   Hence, the gas velocities associated with magnetic activity on the cool star in RS Vul were slightly higher than those achieved by $\beta$ Per.  Moreover, the ($V_x$,$V_y$) slices showed that gas stream was more elongated in RS Vul than in $\beta$ Per, but not as impressive as the distinctive extended flow along the predicted path seen in the 3D image of U CrB.  In both systems, the stream path may have been deflected beyond the central plane by the magnetic field of the cool star.

Both the 3D H$\alpha$ tomography and the VLBI radio images support an earlier prediction of the superhump phenomenon in $\beta$ Per: that the gas between the stars is threaded with a magnetic field even though the hot B8V mass gaining star is not known to have a magnetic field.

\acknowledgements
This research was partially supported by the Russian Foundation for Basic Research (RFBR) grants 09-02-00993 and 12-02-00393, and National Science Foundation grant AST-0908440.   MR is grateful to Alex Cocking for his assistance with the 2D images shown in Figure 1. The VTK software was used in this work to make Figures 7 and 8.

\clearpage

\end{document}